\documentclass{elsart1p}
\usepackage{graphicx}
\usepackage{amssymb}
\begin{document}
\begin{frontmatter}
% notes:
% use the corauthref command within \author for corresponding author
% footnotes;
\title{Jet energy loss in heavy ion collisions \\ from RHIC to LHC energies}

\author{Peter Levai}

\address{KFKI RMKI Research Institute for Particle and Nuclear Physics of HASc, \\
P.O.B. 49, Budapest 1525, Hungary}

\begin{abstract}
The suppression of hadron production originated from the
induced jet energy loss is one of
the most accepted and well understood phenomena in heavy ion 
collisions, which indicates the formation of color deconfined
matter consists of quarks, antiquarks and gluons.
This phenomena has been seen at RHIC energies and now the first
LHC results display a very similar effect. In fact, the suppression
is so close to each other at 200 AGeV and 2.76 ATeV, 
that it is interesting to investigate if such a suppression pattern
can exist at all. We use the Gyulassy-Levai-Vitev description
of induced jet energy loss combined with different nuclear shadowing
functions and describe the experimental data.
We claim that a consistent picture can be obtained for the
produced hot matter with a weak nuclear shadowing. 
The interplay between nuclear shadowing
and jet energy loss playes a crucial role in the understanding
of the experimental data.
\end{abstract}

\end{frontmatter}

% main text
%%%%%%%%%%%%%%%%%%%%%%%%%%%%%%%%%%%%%%%%
\section{Introduction}
\label{sec_intro}
%%%%%%%%%%%%%%%%%%%%%%%%%%%%%%%%%%%%%%%%

Energy loss of high energy quark and gluon jets penetrating  dense
deconfined matter produced in ultrarelativistic heavy ion collisions
leads to jet quenching and thus  probes the quark-gluon plasma formed
in those reactions~\cite{gptw,mgxw92,bdms,bdms8,zahar,urs00,glv2,glv2b}.
The non-abelian radiative energy loss suppresses the hadron production
yield in the momentum range $2-3 \ {\rm GeV/c} < p_T < 15-20 \ {\rm GeV/c}$
at RHIC energies~\cite{PHENIXAuAu,PHENIXdatach,PHENIXdatapi,STARAuAu}.
Recent preliminary experimental data on charged hadron production
collected in PbPb collisions at LHC energies, especially at 2.76 ATeV, 
display the same phenomena~\cite{ALICEdat11}.
Comparing the nuclear modification factor, $R_{AA}$, at RHIC and LHC
energies, one can see that the measured values are very close 
to each other in the momentum range $5 \ {\rm GeV/c} < p_T < 15 \ {\rm GeV/c}$.
This finding is very much surprising, since the measured 
charged hadron multiplicities display a factor of 2
enhancement at LHC energy with respect to the values at RHIC 
energies~\cite{ALICEmult11}.
Thus we expect a two times higher entropy production, which
can be connected to a two times higher density of colored
parton density formed in the central heavy ion collisions.

In order to investigate
the influence of radiative energy loss
on  hadron
production, we apply a perturbative QCD
(pQCD) based description of
heavy ion collisions, including energy loss prior
to hadronization.
First, we check  that the applied pQCD description
reproduces data on charged production
in $p+p$ collision at LHC energies.
Our results are based on a leading order (LO) pQCD analysis.
Detailed discussion of the formalism is
published in Refs.~\cite{Field,PLF00,YZ02}.
We performed next to leading order calculations~\cite{BGG06},
the results will be indicated in $pp$ collisions, however, 
for simplicity we would like to discuss our results at LO
and compare the experimental data.

\begin{figure}[!b]
\centering
\includegraphics[width=0.58\textwidth]{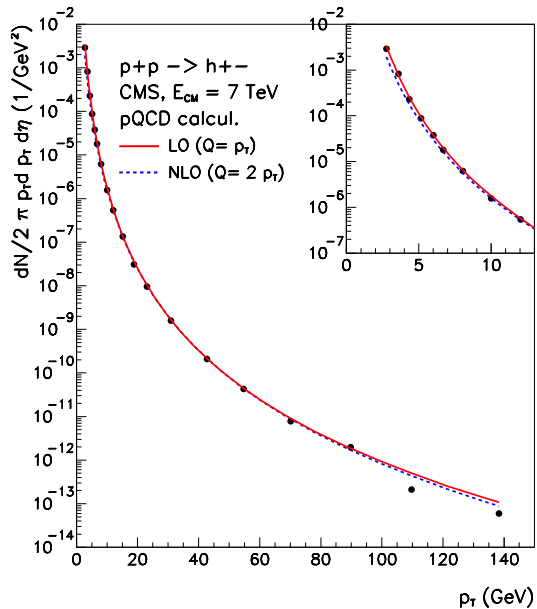}
\caption[]{Charged hadron yield in $pp$-collisions at $\sqrt{s} =$ 7 TeV 
measured by the CMS Collaboration~\cite{CMSdata7}.
The calculation is performed in LO (full line) and NLO (dashed line), 
see details in the text.}
\label{fig_pp}
\end{figure}
%%%%%%%%%%%%%%%%%%%%%%%%%%%%%%%%%%%%%%%%
\section{Parton model for $pp$ collisions}
\label{sec_pp}
%%%%%%%%%%%%%%%%%%%%%%%%%%%%%%%%%%%%%%%%

Our pQCD calculations
follows the general description of charged hadron 
production~\cite{Field,PLF00,YZ02}:
\begin{equation}
\label{fullpp}
E_{h}\frac{d\sigma_h^{pp}}{d^3p} =\!
        \sum_{abcd}\!
        \int\!\!dx_1 dx_2 \
        f_{a/p}(x_1,Q^2) f_{b/p}(x_2,Q^2)\
             \frac{d\sigma}{d{\hat t}}
   \frac{D_{h/c}(z_c,{\widehat Q}^2)}{\pi z_c} \,\,\, .
\end{equation}

Here we use LO parton distribution functions
(PDF) from the MRST98 parameterization \cite{MRS98} and a
LO set of fragmentation functions (FF)~\cite{KKP}.
The applied scales are $Q=\kappa p_T/z_c$ and ${\widehat Q}= \kappa p_T$.
the NLO calculations are performed on the basis of Refs.~\cite{BGG06}.

Utilizing available high transverse-momentum ($5 < p_T < 100$ GeV/c)
$p+p$ data on charged hadron production at $7$ TeV~\cite{CMSdata7}
we can determine the best fitting description.
Fig.~1  shows a general agreement between the data
and our calculations  in LO and in NLO.
The proper scale choice is $\kappa=1$ in LO and
$\kappa= 2$ in NLO calculations.
Interesting to note that we do not need to introduce
intrinsic-$k_T$ at LHC energies, which ingredient
was an essential part of proper
description of hadron production at RHIC energies~\cite{YZ02,BGG06,XNWint}. 
In fact we keep the earlier extracted intrinsic-$k_T$ value at
RHIC energies, but neglect it at LHC energies.

%%%%%%%%%%%%%%%%%%%%%%%%%%%%%%%%%%%%%%%%%%%%%%%%%%%%%%
\section{Nuclear effects in heavy ion collisions}
\label{AA}
%%%%%%%%%%%%%%%%%%%%%%%%%%%%%%%%%%%%%%%%%%%%%%%%%%%%%%
Now we turn to
the calculation for $AA$ collision at RHIC and LHC energies.
We include the isospin asymmetry and the nuclear
modification (shadowing) into the nuclear PDF.
We consider the average nuclear dependence of the PDF,
and apply a scale independent parameterization  with the
shadowing function $S_{a/A}(x)$:
\begin{eqnarray}
f_{a/A}(x) &=& S_{a/A}(x)\left[ \frac{Z}{A} f_{a/p}(x)
+ \left( 1 - \frac{Z}{A} \right) f_{a/n} (x) \right] .
\end{eqnarray}

There are different shadowing functions to be applied, such as the 
relative weak old HIJING-shadowing function~\cite{HIJ} 
and its newer version with and without
$b$-dependence~\cite{HIJnew}. This new version contains a very large 
suppression at low-x, which can be checked by the new LHC results.
Another version, namely the EKS-shadowing function has been developed
with a $Q^2$-dependence but with a weak suppression at low-x, especially
in the high-$Q^2$ domain. In this calculation we will use the EKS99~\cite{EKS99},
which is the basis of the later version and it is a LO-set. 

At RHIC energy we follow our earlier calculations and include
intrinsic-$k_T$ and nuclear multiscattering~\cite{YZ02,XNWint,Papp02},
however, at LHC energies we neglect these contributions.

In our calculation jet quenching reduces  the energy
of the jet before fragmentation. We concentrate on $y_{cm}=0$,
 where the jet transverse
momentum before fragmentation is shifted by the energy loss,
$p_c^*(L/\lambda) = p_c - \Delta E(E,L)$. This
shifts the $z_c$ parameter in the integrand
to $z_c^* = z_c /(1-\Delta E/p_c)$~\cite{WaHu97}.
The applied scale in the FF
is  similarly modified,
${\widehat Q} = \kappa p_T/z_c^*$, while
for the elementary hard reaction the scale
remains $Q=\kappa p_c$.

With these approximations the invariant cross section of hadron
production in central $A+A$ collision is given by~\cite{BGG06}
\begin{eqnarray}
\label{fullaa}
E_{h} \frac{d\sigma_{h}^{AA}}{d^3p} &=&
            \int d^2b \,\, d^2r \,\, t_A(r)
            \,\, t_{A}(|{\bf b} - {\bf r}|) \times \nonumber \cr
        && \sum_{abcd}\!
        \int\!\!dx_1 dx_2 
         f_{a/A}(x_1,Q^2) f_{b/A}(x_2,Q^2)\
             \frac{d\sigma}{d{\hat t}} \frac{z^*_c}{z_c}
   \frac{D_{h/c}(z^*_c,{\widehat Q}^2)}{\pi z_c} \,\,\, .
\end{eqnarray}

Here $t_{A}(b)= \int d z \, \rho_{A}(b,z)$ denotes the nuclear thickness
function,  and it is normalized as usual:
$\int d^2b \, t_{A}(b) = A$. In case of heavy nuclei the Wood-Saxon formula
is applied for the nuclear density distribution, $\rho_{A}(b,z)$.
The integral in $b$ indicates the nuclear overlap in central
collisions and the consideration of the Glauber geometry. 

%%%%%%%%%%%%%%%%%%%%%%%%%%%%%%%%%%%%%%%%%%%%%%%%%%%%%%
\section{Averaged jet energy loss}
\label{QQ}
%%%%%%%%%%%%%%%%%%%%%%%%%%%%%%%%%%%%%%%%%%%%%%%%%%%%%%
Let us summarize our basic knowledge about non-abelian energy loss in
hot dense matter.
First estimates~\cite{gptw,mgxw92} suggested a linear dependence
on the plasma thickness, $L$, namely
$\Delta E \approx 1-2 {\rm \ GeV}(L/{\rm fm})$,
as in abelian electrodynamics.
In BDMS~\cite{bdms,bdms8}, however, non-abelian (radiated gluon final state
interaction) effects were shown to lead to  a quadratic
dependence on $L$ with a larger magnitude of $\Delta E$.
Similar results have been obtained from light-cone path integral
formalism~\cite{zahar,urs00}.

In the GLV formalism, applying opacity series (see Refs.~\cite{glv2,glv2b}),
finite kinematic constraints were found to reduce greatly the energy
loss at moderate jet energies. On the other hand,
the quadratic dependence on $L$ has been recovered in wide energy region.

Non-abelian energy loss in pQCD has been  calculated analytically
in two limits. In the ``thick plasma'' limit,
the mean number of jet scatterings,
$\bar{n}=L/\lambda$, is assumed to be much greater than one.
For asymptotic jet energies the eikonal
approximation  applies and the resummed
energy loss (ignoring kinematic constraints) reduces to the
following simple form~\cite{bdms,bdms8,zahar,urs00}:
\begin{equation}
\Delta E_{BDMS}=\frac{C_{R}\alpha_s}{4}\,
\frac{L^2\mu^2}{\lambda_g} \,\tilde{v} \;\;,
\label{de1}
\end{equation}
where $C_R$ is the color Casimir of the jet ($=N_c$ for gluons),
and $\mu^2/\lambda_g\propto \alpha_s^2\rho$ is a  transport
coefficient of the medium proportional to the parton density, $\rho$.
The factor, $\tilde{v}\sim 1-3$  depends
logarithmically on $L$ and the color Debye screening scale, $\mu$.
It is the radiated gluon mean free path, $\lambda_g$, that enters above.

In the ``thin plasma''  approximation~\cite{glv2,glv2b},
the opacity expansion was applied
and in the first order the following expression was derived
for the energy loss:
\begin{eqnarray}
\Delta E_{GLV}^{(1)}&=& \frac{2 C_R \alpha_s}{\pi}
\frac{E L}{\lambda_g} \, \int_0^1 dx
\int_0^{k_{max}^2} \frac{d {\bf k}^2_\perp}{{\bf k}^2_\perp}
\nonumber \\[.5ex]
&& \hspace{-0.5in}  \int_0^{q_{\max}^2}
\frac{ d^2{\bf q}_{\perp} \, \mu_{eff}^2 }{\pi
({\bf q}_{\perp}^2 + \mu^2)^2 } \cdot
\frac{ 2\,{\bf k}_\perp \cdot {\bf q}_{\perp}
  ({\bf k} - {\bf q})_\perp^2  L^2}
{16x^2E^2 \ + \ ({\bf k} - {\bf q})_\perp^4  L^2 } \;\;.
\label{dnx1}
\end{eqnarray}

\noindent
Here the opacity factor $L/\lambda_g$
is the average number of final state interactions
that the radiated gluons suffer  in the plasma.
The upper transverse kinematic limit is
\begin{equation}
\quad {\bf k}^2_{\max}=\min\, [4E^2x^2,4E^2x(1-x)]\;,
\label{klimits}
\end{equation}
and the upper kinematic bound on the momentum transfer is
 $q^2_{\rm max}= s/4 \simeq 3 E \mu$.

Furthermore,
$\mu_{eff}^2/\mu^2=1+\mu^2/q_{\max}^2$.
At RHIC energies,
these finite limits cannot be ignored~\cite{glv2,glv2b}.
The integral  averages over a screened Yukawa interaction with scale $\mu$.
The integrand is for an  exponential density profile,
$\rho\propto \exp(-2z/L)$, with the same mean thickness, $L/2$,
 as a uniform slab of plasma
of width $L$.

It was shown in Refs.~\cite{glv2,glv2b}
that in the asymptotic $E\rightarrow\infty$
limit, the first-order expression (\ref{dnx1})
reduces to the BDMS result \cite{bdms,bdms8}
up to a logarithmic factor $\log(E/\mu)$.
Numerical solutions revealed that second and third order
in opacity corrections to eq. (\ref{dnx1}) remain  small
$(< 20\, \%)$ in the kinematic range of interest.

\begin{figure}[!t]
\centering
\includegraphics[width=0.49\textwidth]{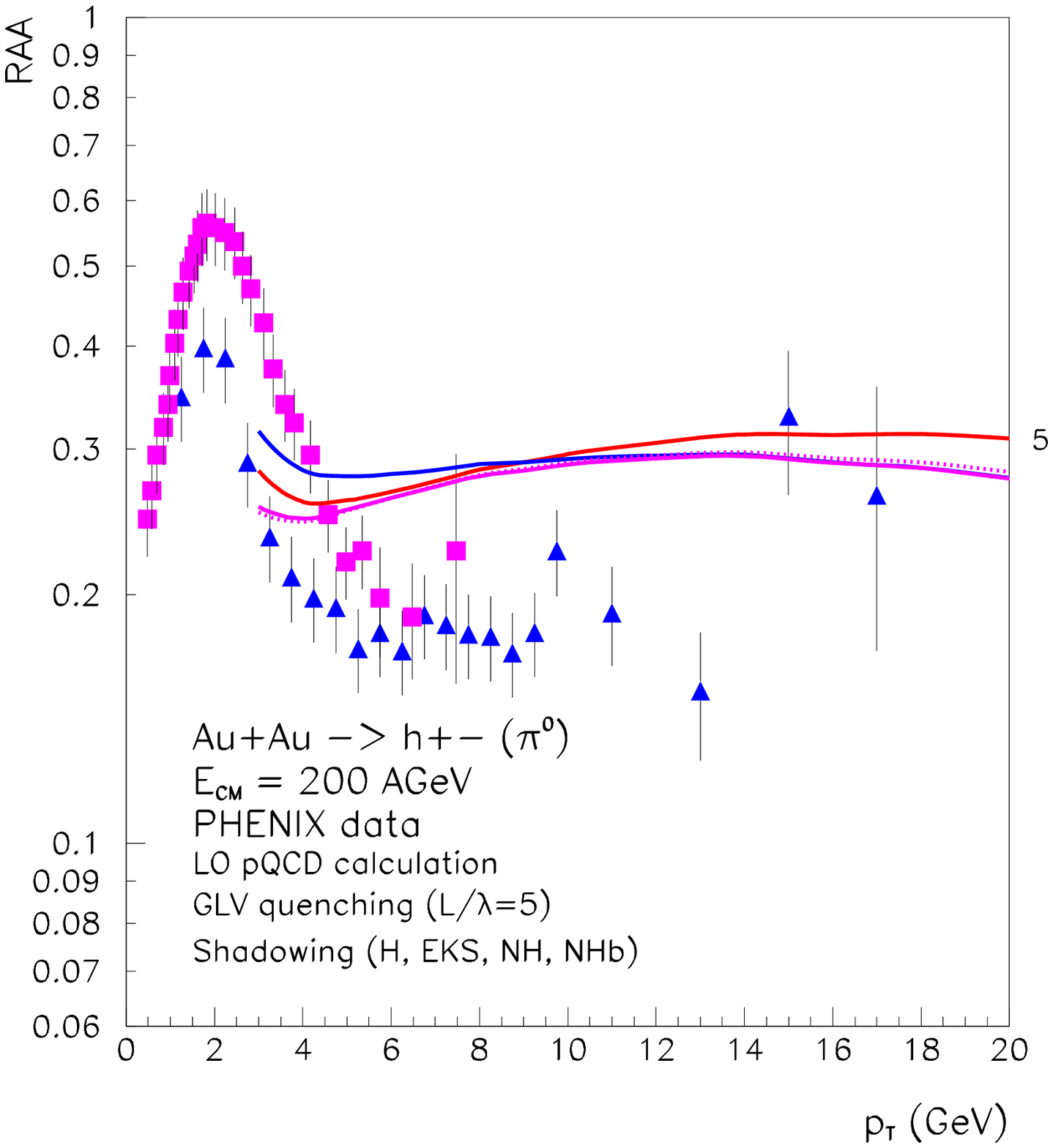}
\includegraphics[width=0.49\textwidth]{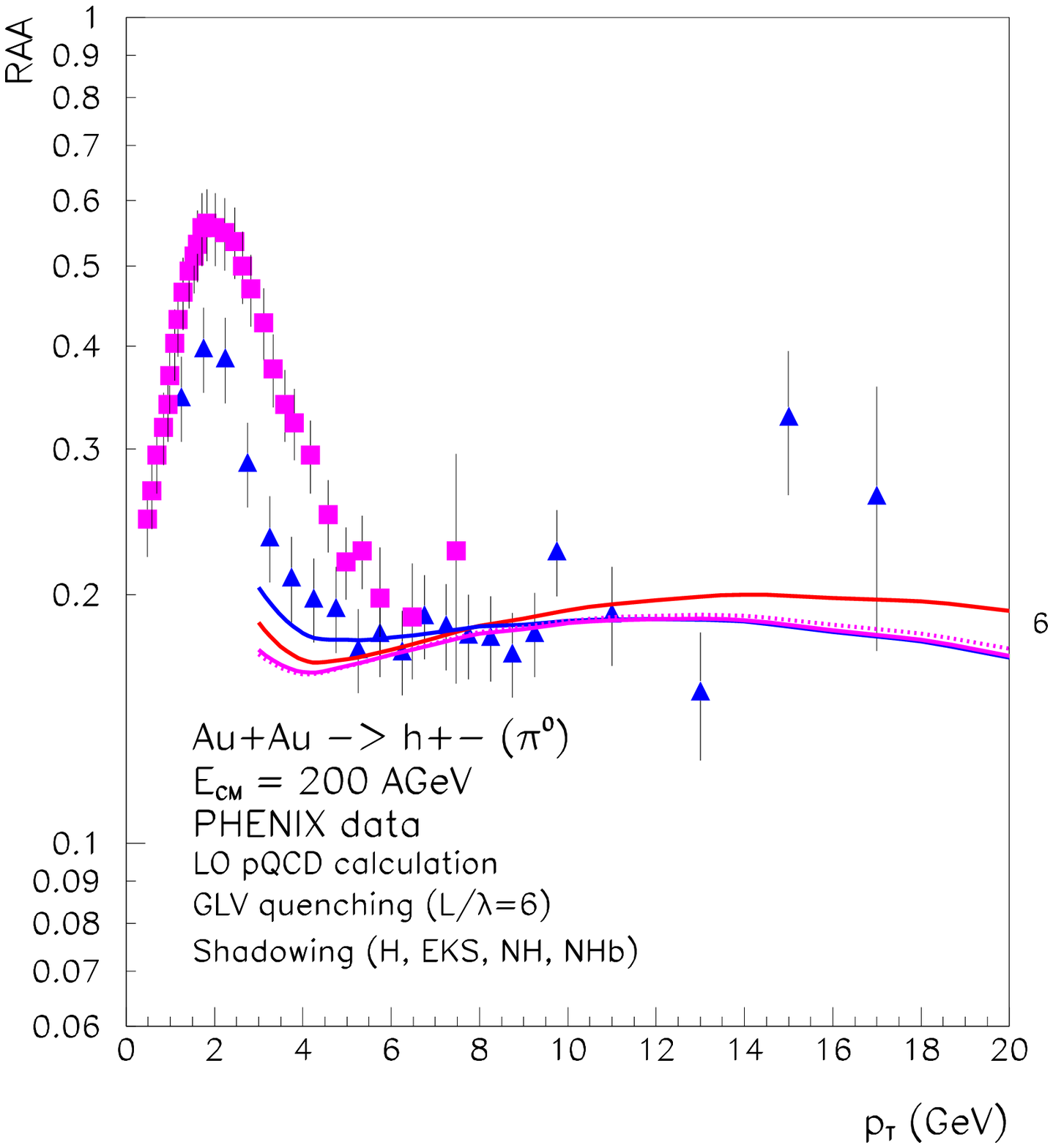}
\caption[]{Calculation of nuclear suppression for charged
hadrons at RHIC energy of $\sqrt{200} \ AGeV$, including
different shadowing functions. 
We choose opacity $L/\lambda = 5$
(left panel) and $L/\lambda = 6$ (right panel).
Experimental data are from PHENIX Collaboration for
charged particle (square)~\cite{PHENIXdatach}
 and $\pi^0$ (triangle)~\cite{PHENIXdatapi}. }
\label{fig_rhic56}
\end{figure}
 
%%%%%%%%%%%%%%%%%%%%%%%%%%%%%%%%%%%%%%%%%%%%%%%%%%%%%%
\section{Numerical results}
\label{num}
%%%%%%%%%%%%%%%%%%%%%%%%%%%%%%%%%%%%%%%%%%%%%%%%%%%%%%
We have calculated jet energy loss at RHIC and LHC energies and
compared our simplified model description to existed data.
We applied the jet quenching effects for a generic plasma
with an average screening scale $\mu=0.5$~GeV, $\alpha_s=0.3$, 
and an average gluon mean free path  $\lambda_g=1$~fm.

In Fig.~\ref{fig_rhic56} we display our results for charged hadrons 
including three types of shadowing
(HIJING, EKS99, new-HIJING and b-dependent new-HIJING) and jet energy loss 
at RHIC energy $\sqrt{s}=200$~AGeV. The opacity which reproduces the 
experimental data is between 5 and 6, we may choose $L/\lambda=5.5$.
This result is consistent with the gluon rapidity densities of
$dN_g/dy = 1000-1400$~\cite{GyulVit02,Vitev05} and the
charged hadron rapidity densities ($\sim 550-650)$~\cite{PHOBOS00,PHENIX01}.
One can see the weak dependence on nuclear shadowing, applying
all three versions.

\begin{figure}[!t]
\centering
\includegraphics[width=0.49\textwidth]{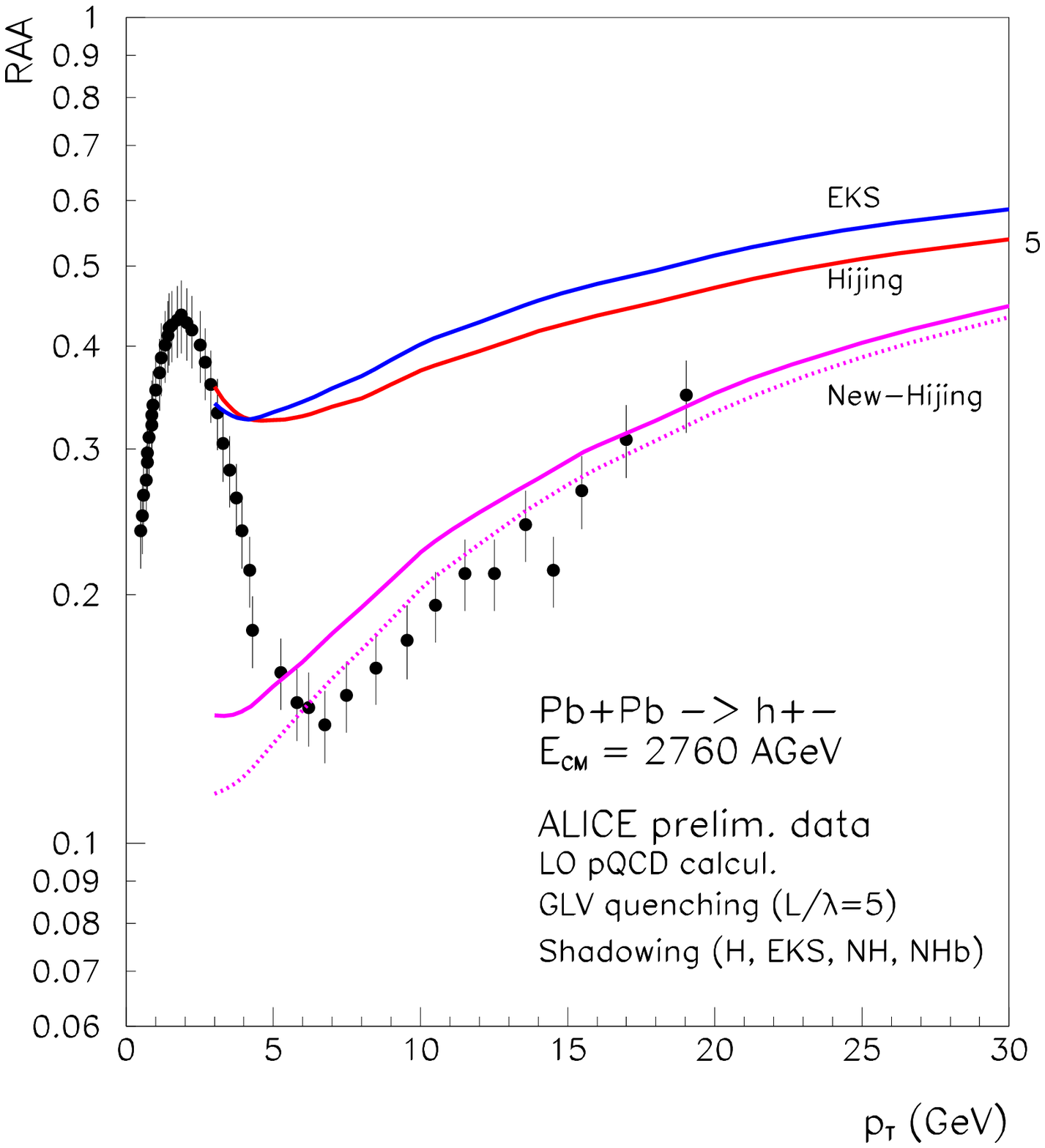}
\includegraphics[width=0.49\textwidth]{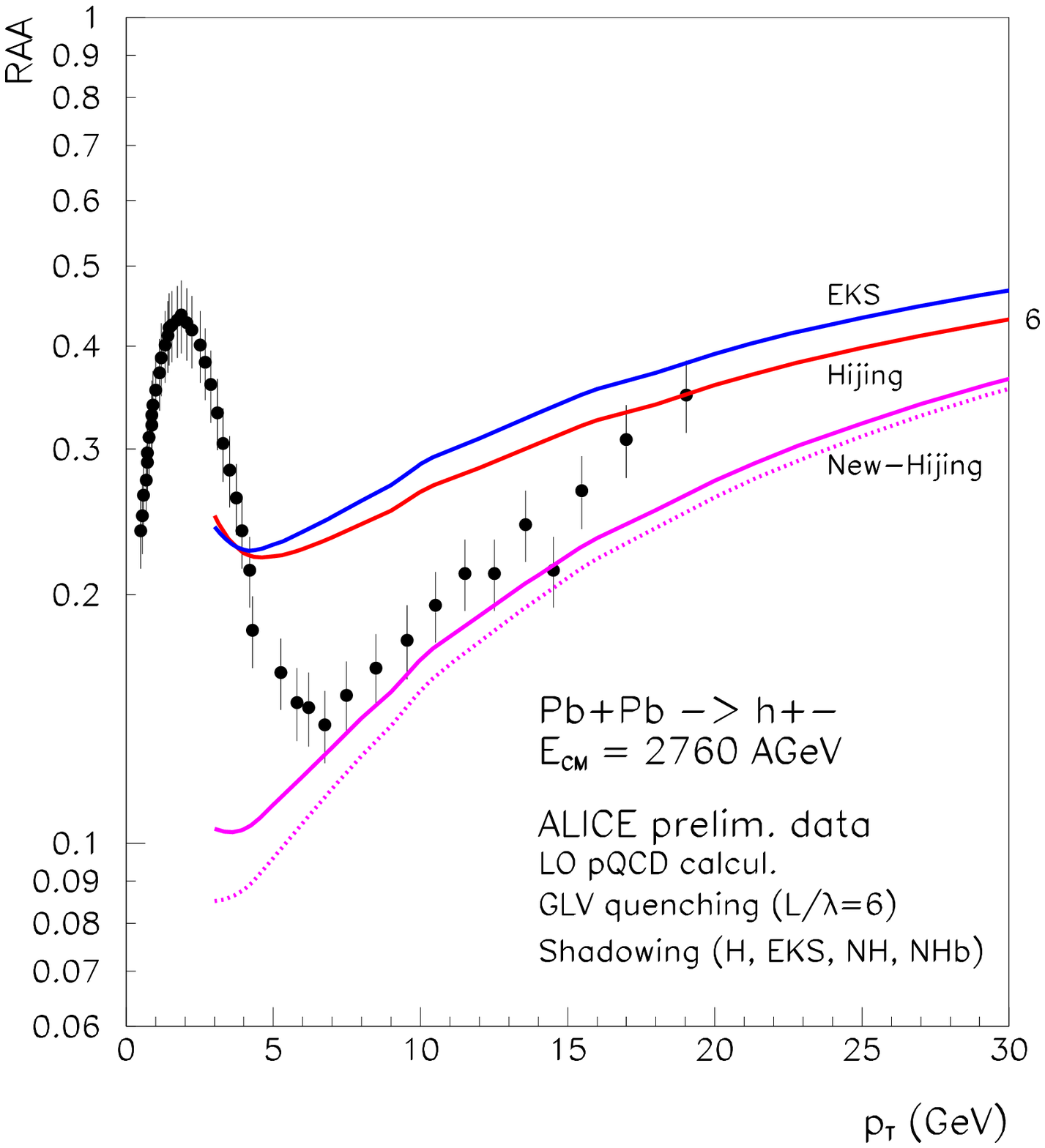}
\caption[]{Calculation of nuclear suppression for charged
hadrons at LHC energy of $\sqrt{s}=2.76 \ ATeV$, including
different shadowing functions. 
We choose opacity $L/\lambda = 5$
(left panel) and $L/\lambda = 6$ (right panel).
Experimental data are from ALICE Collaboration for
charged particle (dots)~\cite{ALICEdat11}. }
\label{fig_lhc56}
\end{figure}

Fig.~\ref{fig_lhc56} displays the results 
in PbPb collisions at $\sqrt{s}=2.76$ ATeV,
applying opacities $L/\lambda=$ 5 and 6. 
We can claim that using the new-HIJING shadowing~\cite{HIJnew}
the requested opacity is also between $L/\lambda=$ 5 and 6,
so $L/\lambda=5.5$ could give also a proper description.

However, this result indicates that both at RHIC and LHC energies
we have the same densities of the colored deconfined matter to be produced.
This result contradicts the total charged multiplicity values
measured at RHIC and LHC energies~\cite{ALICEmult11}.

\begin{figure}[!b]
\centering
\includegraphics[width=0.60\textwidth]{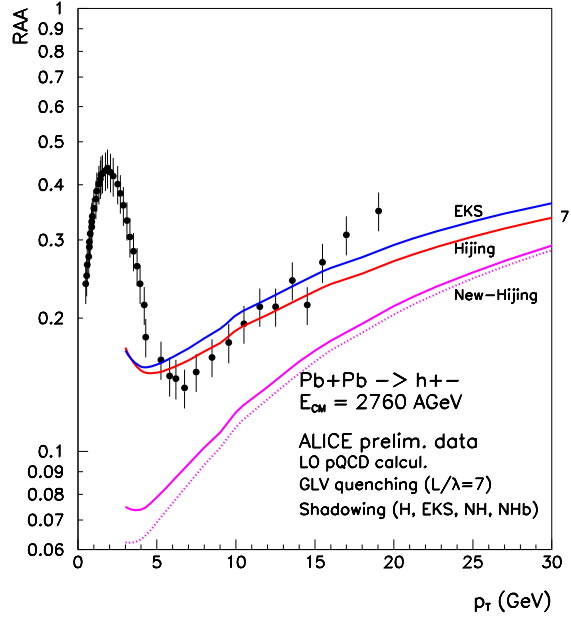}
\caption[]{The nuclear modification factors at 
opacity value $L/\lambda = 7$ at different
shadowing functions, compared to ALICE data at
LHC energy~\cite{ALICEdat11}. The good agreement with the 
old HIJING~\cite{HIJ} and EKS99 shadowing~\cite{EKS99} is emphasized.}
\label{fig_lhc7}
\end{figure}

Thus we increased the opacity and assumed a larger parton densities.
Fig.~\ref{fig_lhc7} displays results, where the
ALICE data are reproduced at $L/\lambda=$ 7, but with the old HIJING
and the EKS99 shadowing.
This means a larger parton density because of the quadratic dependence 
on opacity.

In fact the experimental data on $R_{AA}$ is so close together
at RHIC and LHC, that there is no room for a strong shadowing,
jet energy loss explains the suppression pattern.

Finally, we would like to emphasize the nice agreement between ALICE
data~\cite{ALICEdat11} 
and the jet quenching scenario at high-$p_T$, where the energy loss
becomes weaker and suppression factor approaching unity.
The flat nuclear modification factor seen at RHIC energies disappear.
That flatness was the results of a nice interplay between nuclear shadowing
and multiscattering with the jet quenching. As the flat nuclear modification
factor disappear at LHC energy, then the simple geometrical explanation
of nuclear modification factor also becomes obsolete.
This statement is one of the strongest consequences of the ALICE data.
The other consequence is the negligible role of shadowing at LHC energies.

Our analysis focused on induced jet energy loss and nuclear shadowing.
The introduction of elastic energy loss beyond the radiative energy loss
may change the results, but the basic conclusion on the presence
of a weak nuclear shadowing will not be modified.

Furthermore, these calculations clearly indicate the perturbative QCD region,
where the parton model including jet energy loss is applicable to explain
the measured hadron suppression pattern in heavy ion collisions.
This region is $p_T > 6-7 \ GeV/c$. Under this momentum value the behaviour
of the nuclear modification factor decouples from the pQCD description.
This means that very probably quark coalescence~\cite{HGF1,HGF2,HGF3}
dominates hadron production
in the intermediate momentum region under 6-7 GeV/c, 
both at RHIC and LHC energies, although many questions 
connected to hadron-hadron correlations remained open.

The new ALICE data at $\sqrt{s} = 2.76\ ATeV$~\cite{ALICEdat11}  
induced an extended discussion and many papers appeared
after the publication of these data 
(see e.g.~\cite{New01,New02,New03,New04,New05,New06}).
Final conclusion can be drawn after the new $pp$ data 
collected at  $\sqrt{s} = 2.76\ TeV$ will be analysed and
direct comparison to data with small statistical and systematical
errors can be performed.

%%%%%%%%%%%%%%%%%%%%%%%%%%%%%%%%%%%%%%%%%%%%%%%%%%%%%%
\section{Conclusions}
\label{concl}
%%%%%%%%%%%%%%%%%%%%%%%%%%%%%%%%%%%%%%%%%%%%%%%%%%%%%%
We have investigated jet energy loss at RHIC and LHC energies.
Comparing the experimental data on nuclear modification factor measured in
central Au+Au and Pb+Pb collisions, we can claim the presence of a
weak nuclear shadowing effect to able to accommodate a larger
jet energy loss at LHC energies with respect to the proper opacity values
at RHIC energies. In case of weak nuclear modification factor
the jet energy loss can explain the measured data. 
These calculations should be repeated when experimental data on
reference $pp$-collisions will be available at $\sqrt{s}=2.76 \ $ TeV
and a more precise nuclear modification factor will be displayed.
On the other hand the decreasing tendency of the nuclear modification
factor at $p_T > 8-10$ GeV/c clearly displays the presence of
jet energy loss and these new data at LHC deny the
presence of simple participant nucleon number scaling, which
idea was very popular at RHIC energies to explain the
very flat nuclear modification factor at high-$p_T$
in $AuAu$ collisions . 

\vspace{0.3cm}

\noindent
{\bf Acknowledgment}
I thank the local organizers of ICPAQGP 2010 for a very stimulating conference.
I am indebted to my collaborators on the presented topics, G.G. Barnaf\"oldi,
G. Fai, M. Gyulassy and G. Papp.
This work has been supported by the Hungarian National Science Foundation
OTKA grant no. 77816 and the support of the Hungarian-Indian Bilateral
Agreement of the HASc.

\end{document}